# The DEVStone Metric: Performance Analysis of DEVS Simulation Engines

Román Cárdenas

    Integrated Systems Laboratory, Universidad Politécnica de Madrid, Madrid, Spain, r.cardenas@upm.es
    Dept. of Systems and Computer Engineering, Carleton University, Ottawa, Ontario, Canada

Kevin Henares

    Dpt. of Computer Architecture and Automation, Universidad Complutense de Madrid, Madrid, Spain, khenares@ucm.es

Patricia Arroba

    Integrated Systems Laboratory, Universidad Politécnica de Madrid, Madrid, Spain, p.arroba@upm.es
    Center for Computational Simulation, Universidad Politécnica de Madrid, Madrid, Spain

José L. Risco-Martín

    Dpt. of Computer Architecture and Automation, Universidad Complutense de Madrid, Madrid, Spain, jlrisco@ucm.es
    Center for Computational Simulation, Universidad Politécnica de Madrid, Madrid, Spain

Gabriel A. Wainer

    Dept. of Systems and Computer Engineering, Carleton University, Ottawa, Ontario, Canada, gwainer@sce.carleton.ca

**ABSTRACT**

The DEVStone benchmark allows us to evaluate the performance of discrete-event simulators based on the DEVS formalism. It provides model sets with different characteristics, enabling the analysis of specific issues of simulation engines. However, this heterogeneity hinders the comparison of the results among studies, as the results obtained on each research work depend on the chosen subset of DEVStone models. We define the DEVStone metric based on the DEVStone synthetic benchmark and provide a mechanism for specifying objective ratings for DEVS-based simulators. This metric corresponds to the average number of times that a simulator can execute a selection of 12 DEVStone models in one minute. The variety of the chosen models ensures we measure different particularities provided by DEVStone. The proposed metric allows us to compare various simulators and to assess the impact of new features on their performance. We use the DEVStone metric to compare some popular DEVS-based simulators.

## 1 Introduction

Different Modeling and Simulation (M&S) techniques and tools have been proposed for studying and analyzing human-made or natural systems, each with different options and specific formalism support [1]. Despite efforts to provide compatibility and reusability of models among different M&S tools [2, 3, 4], porting models from one tool to another is still a major challenge that usually implies the complete reimplementation of the models in the new tool.

Among different M&S formalisms, Discrete Event Systems (DESs) are widely used due to their intuitive yet powerful nature [5]. For a given model, these formalisms define a discrete set of states $S$, and how the state of the model changes from $s \in S$ to $s' \in S$ with the occurrence of events. Even though there is a wide variety of DES approaches (e.g., Markov chains or Petri nets), the Discrete Event System Specification (DEVS) formalism [1] stands out as a common denominator for multi-formalism hybrid systems modeling [6]. This feature enables the encapsulation of models described in other formalisms as DEVS models.

A DEVS model is a set of *atomic* and *coupled* models that represent a system hierarchically and modularly. An atomic model specifies the behavior of a system component. The state of an atomic model depends on its previous state and any input event. When an atomic model transitions from one state to another, output events may occur. In contrast, coupled models define how the system components are interconnected. By coupling one model to another, coupled models specify which output events of the former correspond to input events of the latter. DEVS has been applied in a variety of application domains, including decision support systems [7], disease prediction [8], logistic of maintenance operations [9], smart grid infrastructures [10], and traffic analysis [11].

There are multiple DEVS-compliant simulation engines that provide different features to the modelers. Therefore, we need to define comparative methods to decide which of these tools is best suited for our needs. A way to do it is using benchmarking software that measure features of the tools under study (e.g., power consumption or performance) and assign them an unbiased score [12]. The DEVStone benchmarking toolset [13] was introduced with the aim of providing a common method to compare the simulation performance of DEVS simulators.

DEVStone presents four model topologies. Depending on the topology and other configuration parameters, DEVStone generates synthetic DEVS models with different degrees of complexity. The performance of a simulation tool is then measured as the time required to simulate the synthetic DEVStone model. However, state-of-the-art studies select a set of DEVStone models in an arbitrary fashion, making it difficult to compare results of different research works.

Here we present a common evaluation method of DEVS-based simulator performance, a key aspect when deciding the most convenient modeling environment. We extend DEVStone and introduce a common model set to define the *DEVStone metric*. We introduce a basic performance metric to be used as a complete benchmarking technique. We allow the generation of objective and shared ratings reflecting the performance of DEVS-based simulators. This is the benchmark available for comparing the performance of DEVS simulators. The contributions of this work are:

- We revisit the topologies of the original DEVStone benchmarking tool and provide equations to compute model-related parameters (e.g., number of couplings or number of events triggered).
- We define the DEVStone metric as the average number of DEVStone units that a DEVS-compliant simulation engine can simulate in one minute. A DEVStone unit corresponds to a set of 12 DEVStone models with different characteristics (e.g., number of components, interconnections, or simulation events). These models are intended to stress-test the simulators under study. The selected models represent four DEVStone model types with three different topologies each. To avoid too long execution times, we ran multiple models in multiple simulators before selecting the presented DEVStone model set to ensure that the benchmark metric was in the order of minutes in most simulators.
- The DEVStone metric can be represented as a single number that corresponds to the total execution, or as a matrix of the time spent on each of these models. While the former gives us a general idea of the performance of the simulation engine, the latter provides insights of how the structure of the model may affect in the performance of the tool.
- We compute this new DEVStone metric for some of the most popular DEVS simulation engines. We run the benchmark on these tools and discuss the obtained results.

The proposed benchmark metric allows DEVS modelers to compare different available simulation tools and decide which tool is more suitable for them. Furthermore, software developers can integrate this metric in the development process to assess how new features and updates of a simulation tool impacts its simulation performance.

It is worth mentioning that our work focuses on sequential simulation exclusively. Furthermore, we only compare simulators according to the time required to execute the simulations under study. Therefore, we do not consider additional features (e.g., integration with other frameworks or unit testing tools) that modelers might find valuable when deciding which simulator to use. Finally, note that execution times depend on the software and the machine used to run the simulations. Thus, the hardware platform used to replicate the experiments may impact the results, and comparison between simulators should be on the same hardware.

The structure of the paper is organized as follows. First, we introduce some related work in Section 2. In Section 3, we present a detailed description of DEVStone and the proposed benchmarking technique. In Section 4, we use this technique to compare some popular DEVS-based simulators and discuss the results. Finally, we present conclusions in Section 5.

## 2 Related Work

This section first provides a brief introduction to the DEVS formalism and different state-of-the-art DEVS-compliant simulation tools. Finally, we present previous approaches for comparing the performance of these tools.

### 2.1 The DEVS Formalism

DEVS is a formalism for DESs based on set theory [1]. It presents several advantages to analyze and design complex systems (e.g., completeness, verifiability, extensibility, and maintainability). Systems defined using the DEVS theory can be easily implemented using any existing DEVS-compliant computational library. DEVS provides two types of models for defining a system: atomic models, which specify how a system behaves according to an internal state and the occurrence of external events, and coupled models, which defines the structure of the model (i.e., how DEVS subcomponents are interconnected between each other). Chow and Zeigler [14] proposed the Parallel DEVS (PDEVS) formalism, a revision of the original DEVS formalism that enabled the modeling of collisions between internal and external events. It also introduced the concept of message bags, which significantly improved the way modelers could define external transitions when more than one input event happened simultaneously. Currently, PDEVS is the prevalent DEVS variant. In the following, unless it is explicitly noted, the use of DEVS implies PDEVS. Appendix A contains an in-depth description of the DEVS formalism.

### 2.2 DEVS Simulation Engines

Along with PDEVS, Chow and Zeigler defined a formal construct to enable the parallel and distributed execution of PDEVS models. Nowadays, PDEVS is implemented in numerous DEVS simulation engines based on this algorithm. Some of the most popular DEVS simulators are described in the next paragraph.

The adevs simulator [15] is a C++ library for developing models based on the PDEVS and Dynamic DEVS (dynDEVS) formalisms. In terms of performance, adevs is a useful reference, as it usually presents the best results. Cadmium [16] is the latest simulator presented by the ARSLab research group, after CD++ and CDBoost. It is a strongly typed DEVS simulator written in C++ that focuses on ensuring the described model's validity according to the DEVS formalism. It includes a DEVS-based kernel for embedded systems. PyPDEVS [17] is a popular DEVS simulator developed in Python. It has two simulator implementations: a main simulator that allows more configurations, and a minimal version that restricts the simulation functioning to the basics and presents a higher performance. Finally, the xDEVS simulator [18] supports different programming languages: C++, Java, and Python. While all the implementations present a similar API, each version presents different features (e.g., parallel and/or distributed simulation, model flattening, or transducer modules).

## 2.3 Measuring the Performance of DEVS Simulators

Performance of DEVS-compliant simulation engines has been a research topic since the definition of the DEVS formalism. However, the methodologies used to measure the performance speedup of new proposals differ depending on the research. For instance, DEVS-C++ [19] is a high-performance environment that focuses on modeling large-scale systems with high resolution. The authors simulate a watershed with different degrees of detail to compare the speedup of using this tool on different High-Performance Computing (HPC) clusters. Hu and Zeigler [20] proposed an alternative simulation engine for improving the performance of large-scale cellular DEVS models by implementing a data structure that allows less time-consuming searches of active models. The performance analysis consists of a comparison between state-of-the-art simulation engines and the new approach when simulating two example models. Muzy and Nutaro [21] detected that the classical implementation of DEVS simulators could lead to memory inefficiencies resulting from an excessive number of nodes. The authors proposed different simulation algorithms to overcome these deficiencies.

To illustrate the obtained speedup, each of these contributions included a performance comparison between previous simulation engines and simulation engines that implemented the proposed algorithms. These comparisons measured the time required to simulate an arbitrary model with a considerable degree of complexity. However, each contribution used a different model to illustrate this performance enhancement. Depending on the model under study, the number of events, couplings between components, and processing time of the transition functions can vary significantly. Thus, a performance comparative considering only one model is not enough to evaluate a DEVS simulation engine. Additionally, as each contribution used a completely different model to illustrate its performance enhancement, it is not possible to compare the performance enhancement achieved by these contributions.

The DEVStone benchmark [13] was introduced to overcome these limitations. DEVStone describes four types of model structures that can adjust the complexity of the model by depending on four configuration parameters. Section 3 provides an in-depth description of the DEVStone benchmark. Since its introduction, DEVStone has become a 'de-facto' standard for analyzing the performance of DEVS-based simulators. Some authors have used it as a metric to evaluate their DEVS implementations [18, 22]. Others used it for measuring the impact of original proposals for performance improvement [23, 24]. Several works have compared some of the most relevant DEVS-based simulators of the state of the art using this benchmark [25, 26]. Therefore, the DEVStone benchmark can be useful for different purposes. It can help new modelers by giving them a performance insight of the most popular DEVS simulators, facilitating the task of selecting the simulator that fits best for their interests. Also, it can be used for developers to compare their implementations and to evaluate the performance of new proposals and methods for lowering the overhead introduced by the simulation tools.

However, despite the growing popularity of this benchmark, a common metric has not been proposed yet. As DEVStone defines different models and allows to parametrize them to vary their size and complexity, authors must select specific combinations of models and parameters for comparing their implementations. In this way, some of them opt to explore all the combinations in a predefined range. Others establish the reference using a small set of heterogeneous models.

Different DEVStone models variations have also been presented to explore specific simulations aspects. Risco et al. [27] proposed an HOmod model variation called HOmem, that presents a straightforward mathematical way of incrementing the traffic of events with respect to the three simpler DEVStone models. Van Tendeloo and Vangheluwe [26] introduced an HI model variation that removes the recursiveness present in the DEVStone definitions. This alternative model is four atomic fully connected atomic models. Thus, they increment the number of interconnections, and show a quadratic growth in the number of events. In this model, a single parameter is provided for defining whether collisions should happen, being able to measure the bag merging algorithms if this is activated.

This variety of references makes the comparison between works difficult and is contrary to the concept of benchmark as such. This paper aims to solve this heterogeneity of references by defining a common metric for evaluating DEVS-based simulators.

## 3 The DEVStone Metric

In this section, we first introduce the classic DEVStone models. We describe their components and parameters, and their differences are discussed. Then, we select a diverse set of these models to compose the DEVStone metric. This metric is used later in Section 4 as a standard point of reference to compare the most popular state-of-the-art DEVS-compliant simulators.

### 3.1 DEVStone Models

DEVStone [13] enables the generation of multiple synthetic DEVS models with varying shapes and sizes. DEVStone models are simulated to test different features of the simulator under study. DEVStone models are uniquely defined with five parameters:

- *Type*: it defines the number of input/output ports of the model, its structure, and how the model's subcomponents are interconnected with each other.
- *Depth*: it specifies the number of nested models (i.e., levels) in the model hierarchy. The same DEVStone model is repeated iteratively in each of these levels, except for the last one. This inner-most coupled model consists of a single atomic model.
- *Width*: it determines the number of components in each layer of the DEVStone model.
- *Internal transition delay*: it specifies the wall-clock time (in milliseconds) that an atomic model must spend computing its next state when its internal transition function is triggered. When an atomic model's internal transition function is activated, the simulator will run CPU-intensive arithmetic computations for as much time as defined by the internal transition delay, regardless of the DEVS simulator under use. The Dhrystone benchmark [28] is used to emulate CPU-intensive integer arithmetic operations.
- *External transition delay*: this parameter is equivalent to the internal transition delay, but for atomic models' external transition functions. It specifies the wall-clock time (in milliseconds) that an atomic model must spend computing its next state when its external transition function is triggered.

Atomic models have one input port and one output port. When triggered, their output function $\lambda$ produces a message bag with a single integer value, regardless of the number of messages present in their input bag. This measure avoids excessive memory consumption, resulting from the accumulation of the outputs of different atomic models at different coupled models.

At the beginning of the simulation, a single integer value is injected in all the input ports of the topmost coupled model. The simulation time of the model is measured since the introduction of this stimuli until there is no pending events in any atomic model.

The particularities of each DEVStone type are presented below. For each of them, some equations are shown to describe their structure and behavior. Considering the different parameters, they allow to calculate the number of atomic models ($N_{ATOMIC}$), events ($N_{EVENTS}$), External Input Couplings ($N_{EIC}$), External Output Couplings ($N_{EOC}$), and Internal Couplings ($N_{IC}$).

### 3.1.1 LI models

Low Interconnectivity (LI) models present the simplest DEVStone structure, where the only couplings of each depth level are the ones that connect the parent input port with both the coupled and the atomic input ports. There is a single parent output port, connected to the output of the internal coupled model. This composition is depicted in Figure 1.

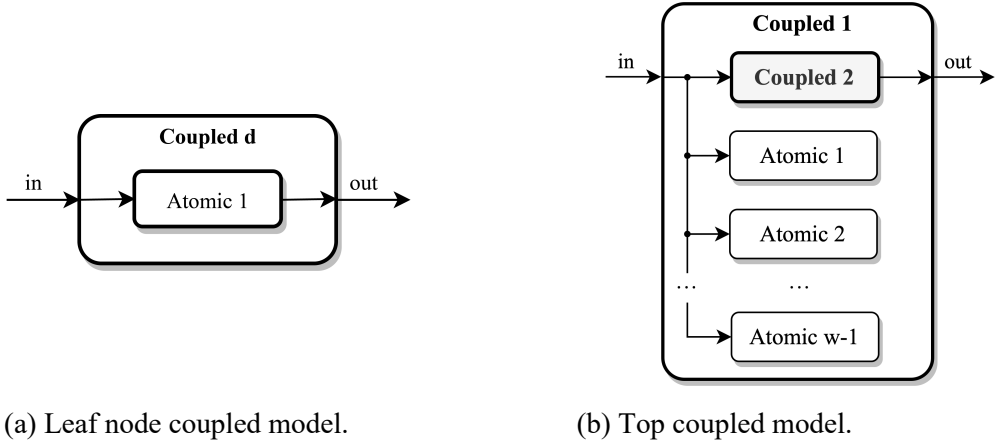

(a) Leaf node coupled model.   (b) Top coupled model.

**Figure 1: Structure of the LI DEVStone model.**

LI models have $d - 1$ layers containing a coupled model and $w - 1$ atomic models, where $d$ is the depth and $w$ is the width of the model. Also, the innermost level contains a single atomic model, as shown in Figure 1(a). Due to the configuration of the couplings, only one internal event and one external event are produced in each atomic model. Among the couplings, the major part corresponds to EIC couplings (there is one for each component, both atomic and coupled). The rest of them correspond to EOC couplings (one per depth level). This model does not present IC couplings. Equation (1) summarizes the LI model's characteristics.

$$\begin{aligned}
N_{ATOMIC} &= (w - 1) \times (d - 1) + 1 \\
N_{EIC} &= w \times (d - 1) + 1 \\
N_{EOC} &= d \\
N_{IC} &= 0 \\
N_{EVENTS} &= (w - 1) \times (d - 1) + 1
\end{aligned} \quad (1)$$

### 3.1.2 HI models

High Interconnectivity (HI) models have the structure shown in Figure 2. They extend the LI model definition adding additional internal couplings between each pair of adjacent atomic models. Figure 2(b) depicts these new internal couplings in gray.

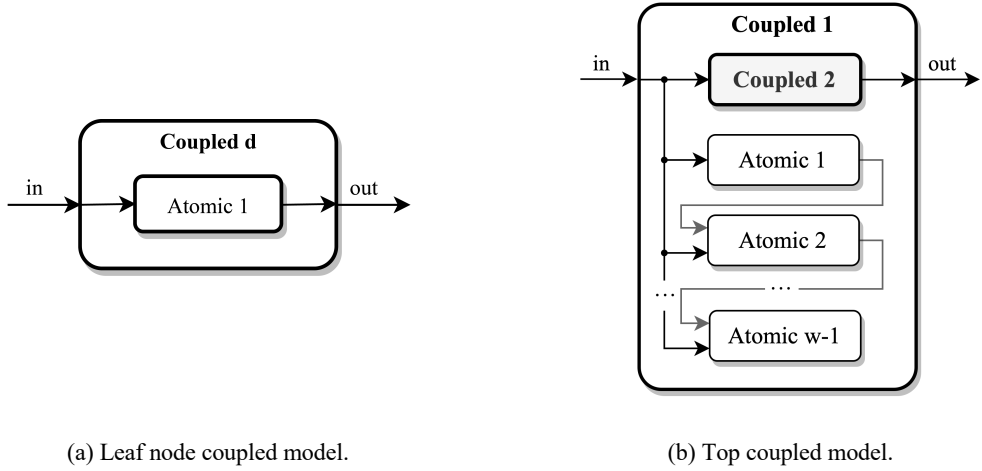

(a) Leaf node coupled model.    (b) Top coupled model.

**Figure 2: Structure of the HI DEVStone model.**

Hence, the number of atomic components, EIC, and EOC couplings remains the same. However, these extensions alter the number of events and IC couplings. In each depth level (except for the last one, that only has an atomic model), $\sum_{i=1}^{w-1} i$ events are produced due to the chaining of the atomic models, as shown in Equation (2). Moreover, there are $w - 2$ IC couplings for each remaining level if $w > 2$. If $w \leq 2$, there are no IC couplings.

$$
\begin{aligned}
N_{ATOMIC} &= (w - 1) \times (d - 1) + 1 \\
N_{EIC} &= w \times (d - 1) + 1 \\
N_{EOC} &= d \\
N_{IC} &= \begin{cases} (w - 2) \times (d - 1), & \text{if } w > 2 \\ 0, & \text{otherwise} \end{cases} \\
N_{EVENTS} &= 1 + (d - 1) \times \frac{(w-1) \times w}{2}
\end{aligned}
\qquad (2)
$$

### 3.1.3 HO models

High Output (HO) models (depicted in Figure 3) extend HI models by adding one output coupling for every atomic model. Also, coupled models have two input and two output ports instead of one.

For a given coupled model, there are two external input couplings connecting its two input ports with the matching input ports of its child coupled model. Additionally, there is one external input coupling between the second input port of the parent coupled model and each of its $w - 1$ child atomic models. There is one external output coupling that interconnects the first output of the child coupled model with the first output of its parent coupled model. On the other hand, the second output port of coupled models remains unconnected. This can help us to detect memory leakage issues of simulation engines that do not clean events of unconnected ports. The overall number of internal couplings, atomic models, and events remains the same as in HI models. Equation (3) shows the characteristics of the HO models.

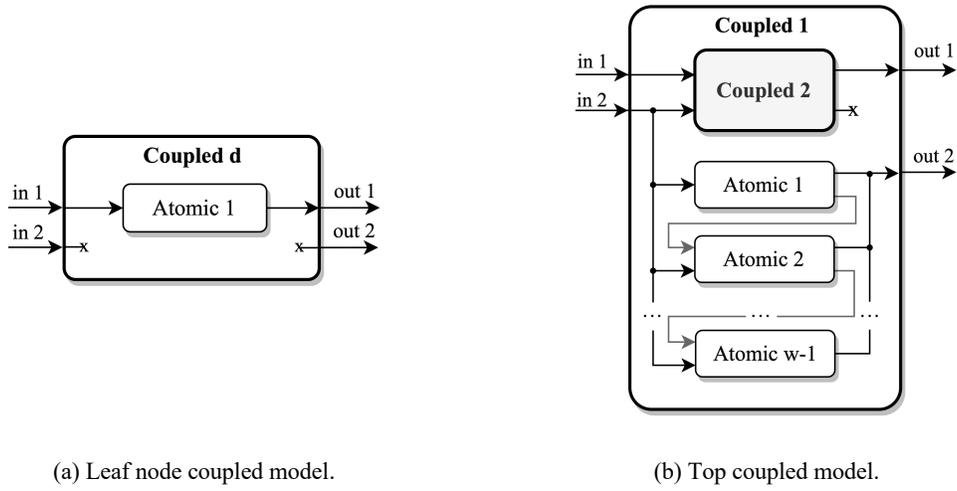

(a) Leaf node coupled model.   (b) Top coupled model.

**Figure 3: Structure of the HO DEVStone model.**

$$\begin{aligned}
N_{ATOMIC} &= (w-1) \times (d-1) + 1 \\
N_{EIC} &= (w+1) \times (d-1) + 1 \\
N_{EOC} &= w \times (d-1) + 1 \\
N_{IC} &= \begin{cases} (w-2) \times (d-1), & \text{if } w > 2 \\ 0, & \text{otherwise} \end{cases} \\
N_{EVENTS} &= 1 + (d-1) \times \frac{(w-1) \times w}{2}
\end{aligned} \quad (3)$$

### 3.1.4 HOmod models

The structure of modified HO models (HOmod) is shown in Figure 4.

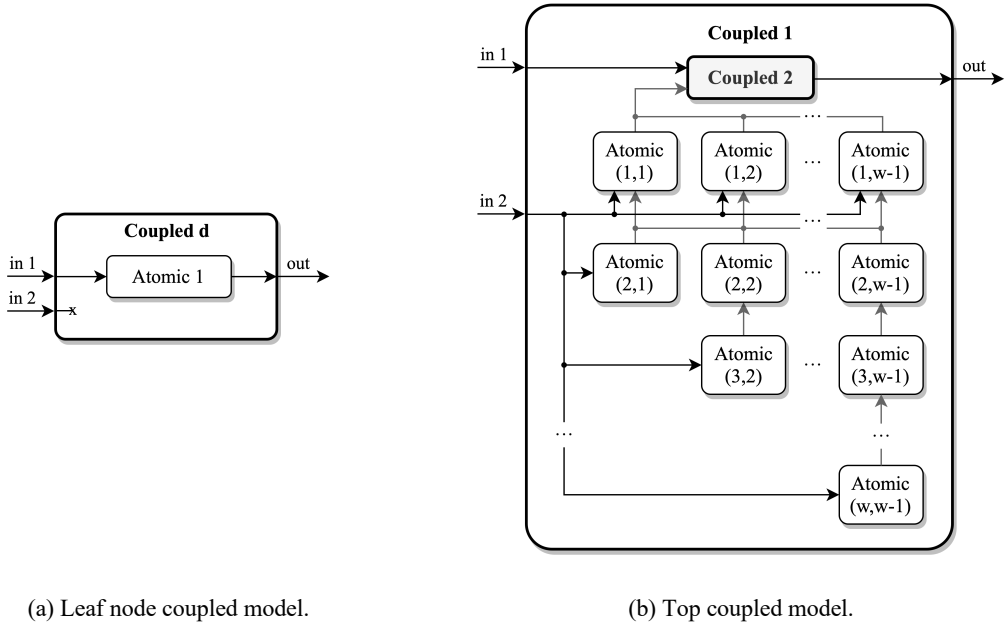

(a) Leaf node coupled model.   (b) Top coupled model.

**Figure 4: Structure of the HOmod DEVStone model.**

In HOmod models, coupled components have two input ports and one output port. As depicted in Figure 4(a), the innermost coupled model only contains one atomic model. However, the remaining $d-1$ coupled components arrange their atomic subcomponents in $w$ rows and $w-1$ columns, as shown in Figure 4(b). The first row has $w-1$ atomic models (i.e., atomic models $(1,1)$ to $(1, w-1)$ in Figure 4(b)). On the other hand, the $w-1$ remaining rows resemble an upper triangular matrix, in which all the elements below the main diagonal are empty and the other elements contain an atomic model. For instance, the second row contains $w-1$ elements (atomic models $(2,1)$ to $(2, w-1)$), while the third row contains $w-2$ elements (atomic models $(3,2)$ to $(3, w-1)$). The number of atomic models per row decreases one by one until row $w$, which has only one atomic model (in Figure 4(b), the atomic model $(w, w-1)$).

The first input port of coupled models is only connected to the first input port of their child coupled model. On the other hand, the second input port is connected to the $w-1$ atomic models of the first row and to the atomic models placed at the main diagonal of the upper triangular matrix comprised by the remaining $w-1$ rows (e.g., atomic models $(2,1)$, $(3,2)$, or $(w, w-1)$).

The output of all the atomic models in the first row is connected to the second input port of the child coupled component. Additionally, the output of every atomic model in the second row is connected to the input of all atomic models in the first row. The output of each atomic model in the remaining $w-2$ rows is only coupled to the input of the atomic model placed in the row above and the same column. For instance, the output of atomic model $(4, w-1)$ is coupled to the input of atomic model $(3, w-1)$, whose output is coupled to atomic model $(2, w-1)$. However, as the atomic model $(2, w-1)$ belongs to the second row, its output is coupled not only to the input of the atomic model $(1, w-1)$, but to all the atomic models in the first row.

The output port of all the coupled models is connected to the output port of their parent coupled model. This configuration results in the following number of atomic models, events and couplings:

$$
\begin{aligned}
N_{ATOMIC} &= \left[w - 1 + \frac{(w-1) \times w}{2}\right] \times (d-1) + 1 \\
N_{EIC} &= [2 \times (w-1) + 1] \times (d-1) + 1 \\
N_{EOC} &= d \\
N_{IC} &= \left[(w-1)^2 + \frac{(w-1) \times w}{2}\right] \times (d-1) \\
N_{EVENTS} &= 1 + \sum_{i=1}^{d-1} \left[(1 + (i-1) \times (w-1)) \times \frac{(w-1) \times w}{2} + (w-1) \times \left(w + (i-1) \times (w-1)\right)\right]
\end{aligned}
\tag{4}
$$

## 3.2 The DEVStone Benchmarking Metric

The *DEVStone benchmarking metric*, $D_{i,j}$, is defined as the average number of *DEVStone units* that a machine $i$ can execute in one minute using the DEVS simulator $j$. A DEVStone unit is defined as the 12 DEVStone models that comprise the *DEVStone benchmark set*, $\mathcal{B}$.

The benchmark set $\mathcal{B}$ intends to stress-test the simulation engines under study using DEVStone models with different characteristics (e.g., couplings, subcomponents, or events). On the other hand, we also wanted the DEVStone metric to be in the order of minutes in most simulators, so running the benchmark would not take too long. Thus, we ran multiple DEVStone models in different simulators before deciding which models would be part of the benchmark set. The selected models are presented in Table 1.

**Table 1: DEVStone metric model set. All the models are configured to have 0 internal and external transition delay.**

| Model Type | Depth | Width |
|---|---|---|
| LI | 200 | 200 |
|  | 200 | 40 |
|  | 40 | 200 |
| HI | 200 | 200 |
|  | 200 | 40 |
|  | 40 | 200 |
| HO | 200 | 200 |
|  | 200 | 40 |
|  | 40 | 200 |
| HOmod | 20 | 20 |
|  | 20 | 4 |
|  | 4 | 20 |

We use three models per DEVStone model type. For the LI, HI, and HO models we use one "*balanced*" model configured to have depth of 200 levels and a width of 200 components per level. Then, we define two additional models: (i) a "*deep*" model (i.e., a model with a depth of 200 levels but a width of only 40 components per level), and (ii) a "*wide*" model whose depth is reduced to 40 levels, but the width remains in 200 components per level. Using models with different shapes allows a detailed study of how the structure of a model may affect the performance of the simulation engine. The three HOmod models follow a similar pattern (i.e., they represent a balanced, a deep, and a wide version of this DEVStone type). However, due to the complexity of these models, using the same configuration for the width and the depth would imply unbearable simulation times even for the fastest simulators. Thus, the depth and width are set to 20 and 20 for the balanced model, 20 and 4 for the deep model, and 4 and 20 for the wide model.

The execution of each one of the models is triggered by inserting a single integer value in all the input ports of their top-most coupled model. It is worth noting that only the simulation time is considered, ignoring the time required for creating the DEVS model and building its corresponding simulators/coordinators hierarchy tree.

In this benchmark, the internal and external transition delays are set to 0. These delays emulate the complexity of computing the next state of the atomic models, which depends strongly on particular use cases, and is not related to the performance of the underlying simulation engine. The proposed DEVStone metric captures the maximal performance differences between the simulators under study. With zero delays, we only measure simulator-related execution times. Otherwise, non-zero transition delays would disguise performance differences between the simulators under study. Non-zero delays might be of interest when working with parallel simulators. These use thread-safe operations, leading to higher execution overheads than sequential simulators. However, parallel simulators can compute state transitions in parallel, outweighing this difference and outperforming sequential simulators. As future work, we will develop an alternative benchmark with non-zero transition delays to compare parallel simulation engines.

The execution time for each model $m \in \mathcal{B}$ using the machine $i$ and the DEVS simulator $j$, $T_{i,j}^m$, is computed as an average over N simulation replications:

$$T_{i,j}^m = \frac{1}{N}\sum_{n=1}^{N} T_{i,j}^{m,n}, \qquad (5)$$

where $T_{i,j}^{m,n}$ is the simulation time (in seconds) for the model $m \in \mathcal{B}$ during the n[th] replication using the machine $i$ and the DEVS simulator $j$. Every model must be executed enough times to provide acceptable confidence bounds. Thus, we propose that N must be greater than or equal to 30 [29]. The *DEVStone metric* is then defined as the number of DEVStone units that a given computer $i$ can execute per minute using the DEVS simulator $j$:

$$D_{i,j} = \frac{60}{\sum_{m \in \mathcal{B}} T_{i,j}^m} \quad [\text{DEVStones / minute}]. \qquad (6)$$

The resulting DEVStone execution times depend on both the implementation details of the DEVS simulation engine and the architecture of the workstation that executes the simulations. Thus, to compare the performance of different DEVS simulators, this metric must be measured in the same machine. Note that $D_{i,j}$ only considers the execution time, ignoring load times.

## 4 Comparison of DEVS Simulators

In this section, we use the benchmarking definition to evaluate the performance of some popular DEVS simulators. For each of them, a DEVStone implementation has been implemented (available in a public repository [30]). Specifically, the following simulators are considered for this comparison: adevs [27], Cadmium [28], PyPDEVS [17] (for both the main and minimal versions using the PyPy Python implementation, as suggested by the authors), and the C++, Java, and Python implementations of xDEVS [18]. For the Python implementation, we show the simulation times for the basic simulation mode and including a recent feature for applying shared-memory techniques in the model ports (the so-called chained simulation algorithm [24]).

All these simulators present a port-based implementation. Therefore, the models include message entry/exit points (ports) that are linked by specifying source-destination links (couplings). Some additional details about the simulators and interpreters/compilers used for executing the simulations are shown in Table 2. The *Events container* column refers to the data structure used by the different simulators to store the set of new messages in the ports (message bag). The *Components container* column refers to the data structure used to store the different child components in the coupled models.

**Table 2: Summary of simulator versions and main data types and used interpreters / compilers.**

| Engine | Version | Language | Interpreter / Compiler | Events container | Components container |
|---|---|---|---|---|---|
| adevs | 3.3 | C++ 17 | g++ 7.5.0 -O3 | array | std::set |
| Cadmium | 0.2.5 | C++ 17 | g++ 7.5.0 -O3 | std::vector | std::vector |
| PyPDEVS | 2.4.1 | Python 3 | Pypy 7.3.1 | dict | list |
| xDEVS (1) | 1.0.0 | C++ 11 | g++ 7.5.0 -O3 | std::list | std::list |
| xDEVS (2) | 1.0.0 | Java 11 | OpenJDK 11.0.7 | LinkedList | LinkedList |
| xDEVS (3) | 1.1 | Python 3 | CPython 3.6.9 | deque | list |

The DEVStone implementations for all these simulators were run in an Ubuntu 18.04, Intel Core i7-9700 and 64 GB RAM workstation. The benchmark was executed sequentially (i.e., using a single processor). The results are shown in Table 3. The second column shows the accumulated simulation time for all the models in the model set. The third column shows the corresponding DEVStones per minute considering the previous average times. Results are shown with a confidence interval of 95% considering that samples follow a T-distribution for 30 samples [29].

Table 3: DEVStone times for several popular DEVS M&S simulators.

| Engine | Seconds / DEVStone | DEVStones / minute |
| --- | --- | --- |
| adevs | 2.831 ± 0.002 | 21.194 ± 0.014 |
| Cadmium | 76.359 ± 0.081 | 0.786 ± 0.001 |
| PyPDEVS | 126.534 ± 0.147 | 0.474 ± 0.001 |
| PyPDEVS (min) | 30.330 ± 0.031 | 1.978 ± 0.002 |
| xDEVS (C++) | 7.763 ± 0.002 | 7.729 ± 0.002 |
| xDEVS (Java) | 6.074 ± 0.020 | 9.879 ± 0.032 |
| xDEVS (Python) | 72.720 ± 0.127 | 0.825 ± 0.001 |
| xDEVS (Python, chained) | 46.838 ± 0.106 | 1.281 ± 0.003 |

The adevs simulator obtained the best results, being able to perform more than 21 DEVStones per minute. The Java and C++ versions of xDEVS were the next fastest simulators, with 9.879 and 7.729 DEVStones per minute, respectively. On the other hand, Cadmium and all the Python simulators with no optimizations ran less than 1 DEVStone per minute. While the performance of the Python simulators seems reasonable (Python is an interpreted language with higher execution overheads compared to C++ or Java), the results for Cadmium may appear surprising. Cadmium was conceived as an educational tool for learning the DEVS formalism. It performs multiple checks throughout the execution of the simulation to ensure that the model strictly follows the DEVS specification (e.g., events are only generated when the $\lambda$ function is triggered). All these checks imply execution overheads that are higher as the complexity of the model increases. The DEVStone times allows us to measure the effect of any optimization in the simulation engine. For example, the minimal version of PyPDEVS was able to run 1.978 DEVStones per minute (i.e., 4.172 times faster than its standard version).

The DEVStone benchmark presented in this research can be used to compare how different model complexity aspects affect the performance of the simulator under study. Figure 5 shows the percentage of execution time spent on each DEVStone model type for all the selected simulators.

The time required for executing the LI models is negligible compared to the other three types for all the simulators except PyPDEVS and the Java implementation of xDEVS. This implies that the simulation algorithms of these simulators potentially present a higher overhead compared to others, reducing the impact of the model complexity on the overall simulation time. In contrast, Cadmium spends 68.16% of the total execution time running the HO and HOmod models (more than any of the other simulators). This indicates that Cadmium's performance is more sensitive to the complexity of the model under study than the rest.

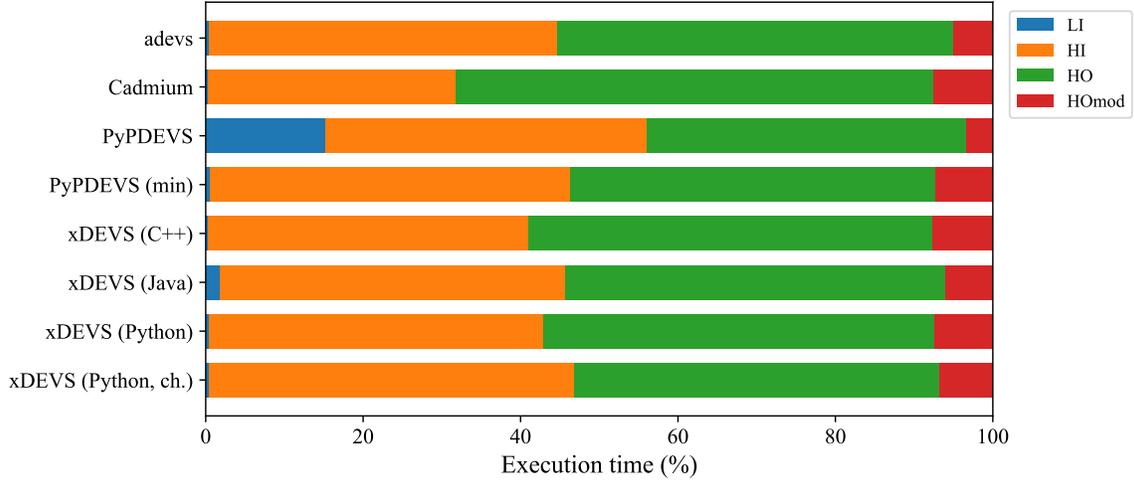

**Figure 5: Percentage of the execution time spent on each DEVStone model type.**

It is possible to expand even more the results obtained running the DEVStone benchmark to see how the width and the depth of the DEVS model impacts on the performance of the simulator. Table 4 displays the execution time for each LI and HI model. There, we can see how the simulators perform for the different model configurations of these DEVStone model types.

**Table 4: Simulation times (in seconds) the LI and HI combinations in the DEVStone benchmark model set. Shapes are provided in the format depth-width.**

| Engine | LI 200-200 | LI 200-40 | LI 40-200 | HI 200-200 | HI 200-40 | HI 40-200 |
| --- | --- | --- | --- | --- | --- | --- |
| adevs | 0.010 ± 0.000 | 0.001 ± 0.000 | 0.001 ± 0.000 | 1.133 ± 0.001 | 0.019 ± 0.000 | 0.099 ± 0.000 |
| Cadmium | 0.149 ± 0.001 | 0.029 ± 0.000 | 0.029 ± 0.000 | 19.843 ± 0.034 | 0.709 ± 0.001 | 3.552 ± 0.019 |
| PyPDEVS | 17.859 ± 0.017 | 0.749 ± 0.002 | 0.730 ± 0.001 | 45.809 ± 0.100 | 1.499 ± 0.006 | 4.263 ± 0.008 |
| PyPDEVS (min) | 0.148 ± 0.000 | 0.020 ± 0.000 | 0.020 ± 0.000 | 11.958 ± 0.015 | 0.330 ± 0.001 | 1.595 ± 0.005 |
| xDEVS (C++) | 0.020 ± 0.000 | 0.003 ± 0.000 | 0.003 ± 0.000 | 2.804 ± 0.004 | 0.055 ± 0.000 | 0.297 ± 0.001 |
| xDEVS (Java) | 0.064 ± 0.002 | 0.023 ± 0.000 | 0.022 ± 0.000 | 2.308 ± 0.015 | 0.098 ± 0.002 | 0.260 ± 0.002 |
| xDEVS (Python) | 0.197 ± 0.001 | 0.038 ± 0.000 | 0.036 ± 0.000 | 25.428 ± 0.113 | 0.954 ± 0.002 | 4.561 ± 0.013 |
| xDEVS (Python, chained) | 0.159 ± 0.001 | 0.029 ± 0.000 | 0.029 ± 0.000 | 17.877 ± 0.067 | 0.644 ± 0.003 | 3.184 ± 0.013 |

The adevs simulator showed the best performance in all the included model configurations, followed by the C++ implementation of xDEVS in most of the DEVStone configurations. On the other hand, the base PyPDEVS simulator was the simulation engine that required more time for simulating the LI models, being up to 90.655 times slower than the Python implementation of xDEVS (second slowest simulator for LI models) and up to 120.669 times slower than the minimal implementation of PyPDEVS. Also, due to the simple structure of LI models, all the simulators listed here obtained similar times for the deep and wide models.

In contrast, the increased complexity of the HI DEVStone significantly impacts on the simulation time. In HI models, wide models require higher simulation times than deep models due to the additional internal couplings (and, therefore, event propagations). In the balanced configuration of HI models, the Java implementation of xDEVS can outperform its C++ counterpart. At the other end, the simulators with worst performance are again the base Python implementations (PyPDEVS and Python xDEVS), with a reduced time difference for this model. Cadmium is the simulator with a more significant simulation time increase comparing LI and HI models. This confirms that Cadmium is the most sensitive to the complexity of the running model.

Table 5 shows the execution time for the models with HO and HOmod structures. There, we can see how the simulators perform for the most complex model configurations that conform the DEVStone benchmark.

**Table 5: Simulation times (in seconds) for the HO and HOmod combinations in the DEVStone benchmark model set. Shapes are provided in the format depth-width.**

| Engine | HO 200-200 | HO 200-40 | HO 40-200 | HOmod 20-20 | HOmod 20-4 | HOmod 4-20 |
|---|---|---|---|---|---|---|
| adevs | 1.288 ± 0.001 | 0.022 ± 0.000 | 0.116 ± 0.000 | 0.138 ± 0.000 | 0.000 ± 0.000 | 0.002 ± 0.000 |
| Cadmium | 38.250 ± 0.055 | 1.023 ± 0.003 | 7.077 ± 0.014 | 5.548 ± 0.008 | 0.034 ± 0.000 | 0.114 ± 0.000 |
| PyPDEVS | 45.679 ± 0.107 | 1.500 ± 0.003 | 4.276 ± 0.013 | 4.063 ± 0.011 | 0.027 ± 0.000 | 0.079 ± 0.000 |
| PyPDEVS (min) | 12.115 ± 0.025 | 0.330 ± 0.001 | 1.598 ± 0.005 | 2.164 ± 0.009 | 0.012 ± 0.000 | 0.040 ± 0.000 |
| xDEVS (C++) | 3.547 ± 0.005 | 0.066 ± 0.000 | 0.370 ± 0.001 | 0.584 ± 0.001 | 0.003 ± 0.000 | 0.011 ± 0.000 |
| xDEVS (Java) | 2.532 ± 0.015 | 0.104 ± 0.001 | 0.298 ± 0.002 | 0.313 ± 0.002 | 0.021 ± 0.000 | 0.031 ± 0.000 |
| xDEVS (Python) | 29.609 ± 0.066 | 1.129 ± 0.003 | 5.417 ± 0.015 | 5.184 ± 0.021 | 0.045 ± 0.000 | 0.121 ± 0.000 |
| xDEVS (Python, chained) | 17.906 ± 0.067 | 0.642 ± 0.003 | 3.207 ± 0.020 | 3.063 ± 0.012 | 0.025 ± 0.000 | 0.071 ± 0.000 |

Again, the adevs simulator outperformed the other simulators in all the included model configurations. However, the Java version of xDEVS obtained better results than its C++ counterpart for all the configurations except the deep configurations of the HO and HOmod model types (i.e., the models with less internal couplings). In HO and HOmod models, consequently with the specification, wide models also require higher simulation times compared to their deep peer.

The increase in complexity of these models also leads to poorer performance results for Cadmium, getting times comparable with the ones obtained by PyPDEVS and the Python versions of xDEVS.

Figure 6 compares the execution time of the selected DEVS-compliant simulation engines for the balanced models that comprise the DEVStone benchmark. For sake of simplicity, we only show the results of the optimized versions of the PyPDEVS and the Python xDEVS simulators. The adevs simulator obtained the best results for all these configurations.

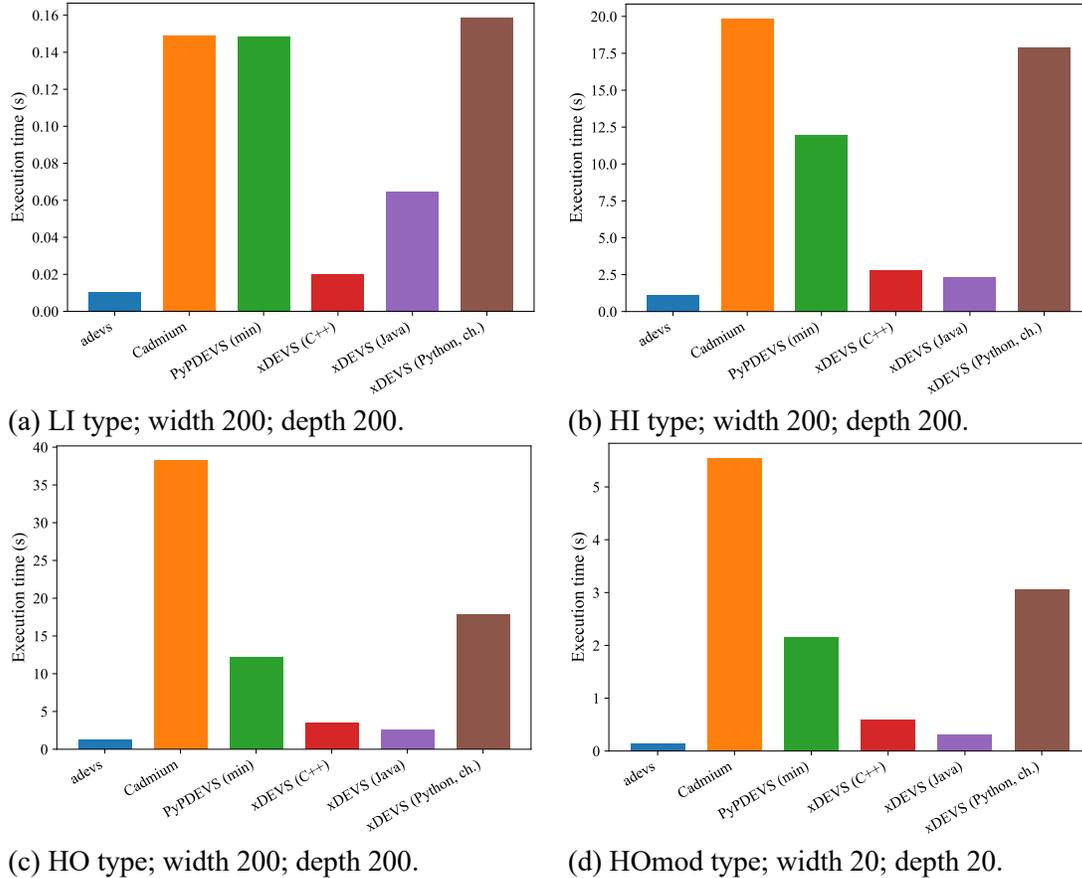

(a) LI type; width 200; depth 200.  (b) HI type; width 200; depth 200.

(c) HO type; width 200; depth 200.  (d) HOmod type; width 20; depth 20.

**Figure 6: Simulation time of DEVS simulators for the balanced models that comprise the DEVStone benchmark.**

Figure 6(a) depicts the execution time for the LI balanced model. Cadmium and the minimal version of PyPDEVS obtained similar results. However, as the complexity of the simulated model increases, the performance of Cadmium decreases. For example, for simulating the balanced HI model (see Figure 6(b)), Cadmium required 19.843 seconds, whereas the minimal version of PyPDEVS finished in 11.958 seconds. In fact, Cadmium ends up being the slowest simulator for the HO and HOmod balanced models (Figure 6(c) and Figure 6(d), respectively). Note that this performance downgrade is mainly due to the multiple runtime checks performed by Cadmium to ensure that the model strictly follows the DEVS formalism. The C++ version of xDEVS obtained the second-best result for the LI balanced model. In contrast, for the rest of the balanced configurations, its Java counterpart outperforms it, showing a better resilience to model complexity.

Figure 7 depicts the mean wall-clock time required by the simulators understudy to execute the deep models comprising the DEVStone model set.

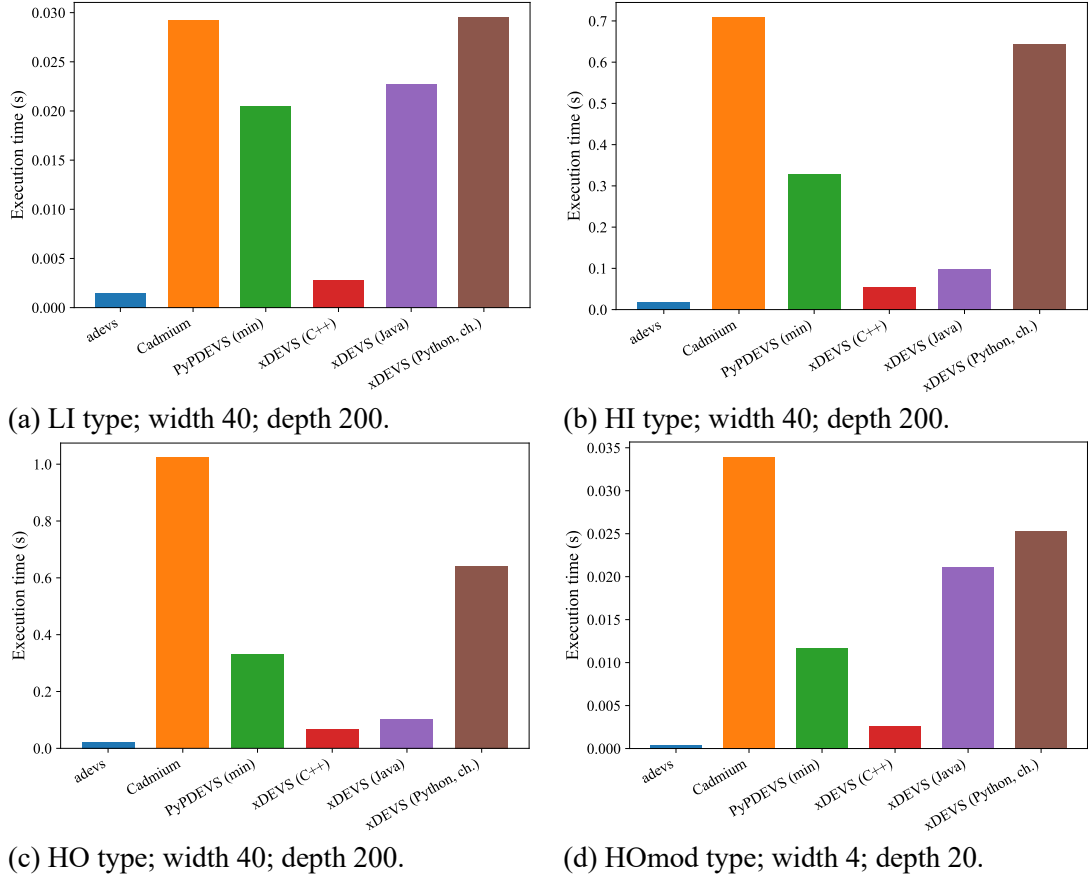

Figure 7: Simulation time of DEVS simulators for the deep models that comprise the DEVStone benchmark.

Once more, adevs outperforms the rest of the simulators in all the deep models. As a rule, the C++ and Java implementations of xDEVS show the second and third best results, respectively. The minimal implementation of PyPDEVS, Python xDEVS, and Cadmium obtained the worst results. However, it is worth mentioning that, for the LI and HOmod models (see Figure 7(a) and Figure 7(d), respectively), the minimal implementation of PyPDEVS achieves better results than Java xDEVS. Note that these are the simplest models in the DEVStone benchmark model set, which indicates a slightly higher computation overhead in the Java version of xDEVS. This overhead becomes negligible as the model complexity increases, requiring more event propagation and model synchronization actions.

Figure 8 displays the execution time of the wide models in the DEVStone model set. If we compare the results of executing the deep and wide LI models (i.e., Figure 7(a) and Figure 8(a)), the results are practically identical. In contrast, for the HI, HO, and HOmod types, wide models take longer to execute than the deep ones, regardless of the simulator under study.

As deep models are more complex than wide models, besides Cadmium, the minimal implementation of PyPDEVS manages to outperform only the Java version of xDEVS (see Figure 8(a)). Furthermore, xDEVS Java shows better results than its C++ analogous for the HI and HOmod deep models (see Figures 8(b) and 8(c), respectively). Again, this proves that the Java implementation shows greater resilience than xDEVS C++.

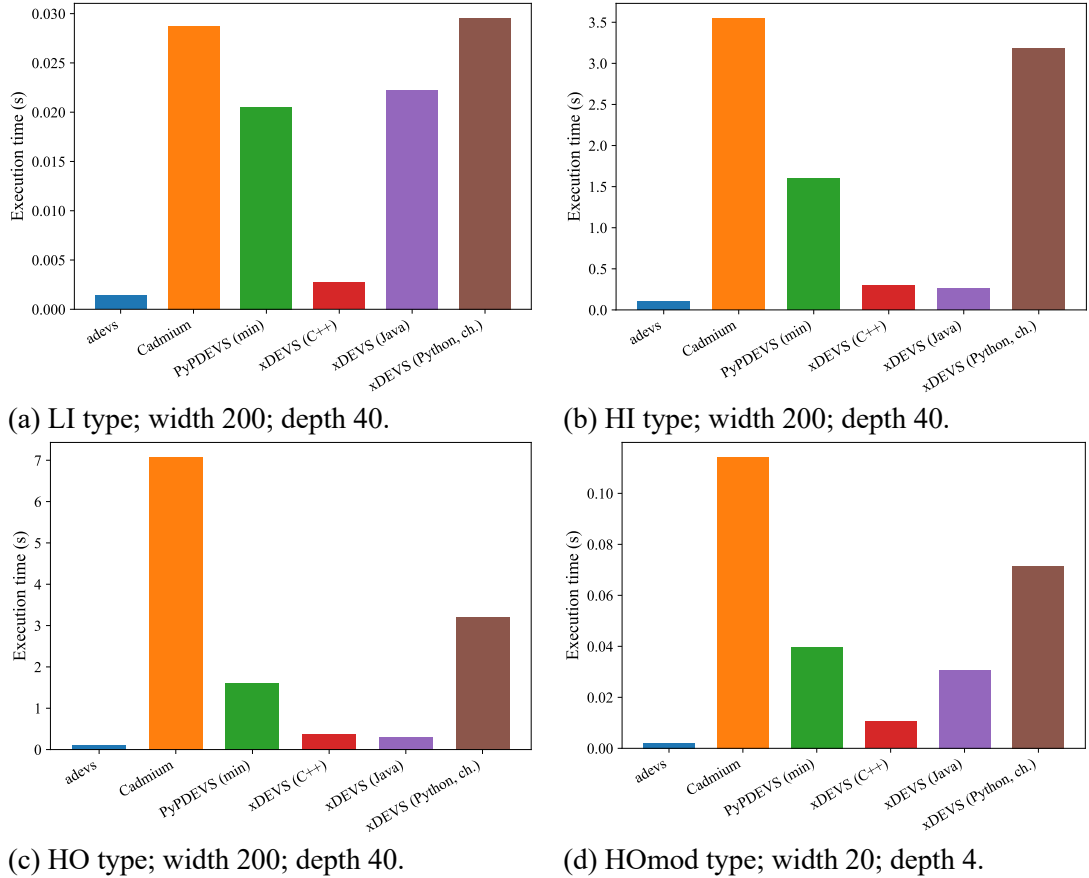

(a) LI type; width 200; depth 40.   (b) HI type; width 200; depth 40.

(c) HO type; width 200; depth 40.   (d) HOmod type; width 20; depth 4.

**Figure 8: Simulation time of DEVS simulators for the wide models that comprise the DEVStone benchmark.**

It is important to remark that a simulation performance comparison is not enough to decide which DEVS simulation tool suits better for a new project. We also must consider additional features provided by each simulator, as these can be very convenient depending on the use case. For example, adevs supports using QEMU [31], Cadmium integrates a DEVS-based Real-Time (RT) kernel for embedded systems [32], and xDEVS provides tools for model unit testing [33].

## 5 Conclusion

The DEVStone benchmark facilitates the process of evaluating the performance of DEVS-based simulators. It allows defining four customizable models, each of them with valuable particularities that allow us to measure specific issues of the simulation. However, this non-homogeneous selection made it difficult to compare the results among studies and conditioned the results according to the chosen models.

The DEVStone metric allows us generating individual and objective ratings for specific pairs of DEVS-based simulators and workstations. The DEVStone metric adds of the simulation times of a fixed mix of synthetic models, which includes a variety of DEVStone models, composed by a selection of balanced, wide, and multi-layer models. This metric was used to compare some popular DEVS-based simulators. Both the global DEVStones/minutes and the specific simulation times per each model have been shown, highlighting the strengths and weaknesses of the different simulators. These results can be used by the authors to enhance the performance of their implementations.

All the DEVStone implementations, for each one of the simulators considered in this article, are available in a public repository [30]. This repository is prepared so that all the simulations can be executed easily from a main script. As future work, we will extend the set of supported simulators present in this repository to cover more popular DEVS-based simulators. Additionally, we will define an alternative DEVStone benchmark to compare parallel and distributed DEVS simulation engines considering the internal and external transition delays of the DEVStone models that comprise this benchmark.

## ACKNOWLEDGMENTS

This project has been partially supported by the Spanish Ministry of Science and Innovation under research grant PID2019-110866RB-I00.

# APPENDICES

## A  The Parallel DEVS Formalism

The Discrete EVent System (DEVS) specification formalism provides a rigorous definition for discrete modeling and simulation. DEVS allows the user to define a mathematical object (i.e., system) that represents an abstraction of real objects. Parallel DEVS (PDEVS) is a popular variant of the original formalism, which addresses some deficiencies of the original DEVS. In fact, multiple state-of-the-art works usually refer to PDEVS as simply DEVS. In this appendix, we provide a brief definition of the PDEVS formalism.

In PDEVS, the behavior of a system can be described at two levels: atomic models, which describe the autonomous behavior of a system as a series of transitions between states and its reactions to external events, and coupled models, which describe a system as the interconnection of coupled components. The formal definition of an atomic model is described as the following:

$$A = \langle X, Y, S, \delta_{int}, \delta_{ext}, \delta_{con}, \lambda, ta \rangle,$$

where:
- $X$ is the set of input events. Each element $x \in X$ corresponds to a possible input event that may trigger the atomic model's external transition function.
- $Y$ is the set of output events. Each element $y \in Y$ corresponds to a possible output event that may be triggered by the atomic model's output function.
- $S$ is the states set. At any given time, the atomic model is in the state $s \in S$.
- $ta: S \to \mathbb{R}_{\geq 0} \cup \infty$ is the time advance function. When the atomic model enters the state $s$, it will remain in this state for $ta(s)$ time units or until the model receives an input event.
- $\delta_{int}: S \to S$ is the internal transition function. After spending $ta(s)$ time units in the state $s$ without receiving any input event, the atomic model transitions to the state $s' = \delta_{int}(s)$.
- $\lambda: S \to Y$ is the output function. When the atomic is about to change its state due to an internal transition, the atomic model generates $\lambda(s) \subseteq Y$ output events. This function is triggered right before calling the internal transition function.
- $\delta_{ext}: S \times \mathbb{R}_{\geq 0} \times X \to S$ is the external transition function. It is automatically triggered when the atomic model receives a bag of input events $X^b \subseteq X$ after $e$ time units since the atomic model transitioned to its current state $s$ (i.e., $0 \leq e \leq ta(s)$). When triggered, $\delta_{ext}(s, e, X^b)$ determines the new state of the atomic model.
- $\delta_{con}: S \times \mathbb{R}_{\geq 0} \times X \to S$ is the confluent transition function. This transition function decides the next state in cases of collision between external and internal events (i.e., $e = ta(s)$). Typically, $\delta_{con}(s, ta(s), X^b) = \delta_{ext}(\delta_{int}(s), 0, X^b)$. The classic DEVS formalism does not provide any mechanism to properly handle state transition collisions. Therefore, the confluent transition function is the main difference between classic DEVS and PDEVS.

On the other hand, the formal definition of a coupled model is described as follows:

$$M = \langle X, Y, C, EIC, EOC, IC \rangle,$$

where:
- $X$ is the set of input events. Each element $x \in X$ corresponds to a possible input event that may enter to the coupled model.
- $Y$ is the set of output events. Each element $y \in Y$ corresponds to a possible output event that may leave the coupled model.
- $C$ is the set of components. Each element $c \in C$ corresponds to a PDEVS submodel (atomic or coupled) of the coupled model.
- $EIC$ is the external input coupling relation. It defines how external inputs of $M$ are linked to component inputs of $C$.

- $EOC$ is the external output coupling relation. It defines how component outputs of $C$ are linked to external outputs of M.
- $IC$ is the internal coupling relation, from component outputs of $c_i \in C$ to component inputs of $c_j \in C$.

Given the recursive definition of M, a coupled model can itself be a part of a component in a larger coupled model, making it possible to construct a hierarchical PDEVS model. Its hierarchical nature eases the reuse, adaptation, verification, and validation of PDEVS models. This definition makes it capable of representing a wide class of other dynamic systems, like differential equations or different types of automata.